\documentclass[final,5p,times,twocolumn]{elsarticle}
\usepackage{latexsym,graphicx,amssymb,amsmath,mathrsfs}
\usepackage{setspace,bm}
\usepackage[colorlinks=true, pdfstartview=FitV, linkcolor=red, citecolor=blue, urlcolor=blue]{hyperref}
\usepackage[caption=false]{subfig}
\usepackage{amsmath}
\usepackage[usenames]{color}
\usepackage{latexsym}

\newcommand\beq{\begin{equation}}
\newcommand\eeq{\end{equation}}
\newcommand{\Slash}[1]{{\ooalign{\hfil/\hfil\crcr$#1$}}}
\newcommand\nc{N_\mathrm{c}}

\newcommand\muiso{\mu_\mathrm{iso}}
\newcommand\mur{\mu_\mathrm{R}}
\newcommand\mui{\mu_\mathrm{I}}
\usepackage[normalem]{ulem}  

\newcommand{\comment}[1]{}

\renewcommand\sout{\bgroup \color{red} \ULdepth=-.5ex \ULset}

\usepackage{amssymb}
\usepackage{amsmath}

\journal{Physics Letters B}

\begin{document}

\begin{frontmatter}



\title{Topological deconfinement transition in QCD at finite isospin
 density\tnoteref{Report}}
\tnotetext[Report]{Report number: YITP-17-04}

\author[YITP]{Kouji Kashiwa}
\author[YITP]{Akira Ohnishi}

\address[YITP]{Yukawa Institute for Theoretical Physics, Kyoto University,
Kyoto 606-8502, Japan}

\begin{abstract}
The confinement-deconfinement transition is discussed from
 topological viewpoints. The topological change of the system is
 achieved by introducing the dimensionless imaginary chemical potential
 ($\theta$). Then, the non-trivial free-energy degeneracy becomes the
 signal of the deconfinement transition and it can be visualized by
 using the map of the thermodynamic quantities to the circle $S^1$ along
 $\theta$. To understand this ``topological'' deconfinement transition
 at finite real quark chemical potential ($\mur$), we consider the
 isospin chemical potential ($\muiso$) in the effective model of QCD.
 The phase diagram at finite $\muiso$ is identical with that at finite $\mur$
 outside of the pion-condensed phase at least in the large-$\nc$ limit
 via the well-known orbifold equivalence. In the present effective
 model, the topological deconfinement transition does not show a
 significant dependence on $\muiso$ and then we can expect that this
 tendency also appears at small $\mur$. Also, the chiral transition and
 the topological deconfinement transition seems to be weakly
 correlated. If we will access lattice QCD data for the temperature
 dependence of the quark number density at finite $\muiso$ with
 $\theta=\pi/3$, our surmise can be judged.

\end{abstract}


\end{frontmatter}



\section{Introduction}
\label{Sec:Intro}

Understanding the confinement-deconfinement transition is one of
the important subjects in the particle and nuclear physics.
In the heavy quark mass limit, the Polyakov loop respecting the
gauge invariant holonomy characterizes the deconfinement transition.
In that case, the spontaneous $\mathbb{Z}_3$ symmetry breaking describes
the deconfinement transition and the Polyakov-loop becomes the
order parameter of the symmetry breaking.
On the other hand, the Polyakov loop cannot be considered as the
order-parameter of the deconfinement transition at finite
quark mass.

Recently, it was found that the confined and deconfined states at zero
temperature ($T=0$) are
characterized by the ground-state degeneracy when the system
has non-trivial topology~\cite{Sato:2007xc}.
This argument is based on the topological order discussed
in the condensed matter physics
where the spontaneous symmetry breaking is absent~\cite{Wen:1989iv}.
Thus, we can expect that the deconfinement transition can be described
by the topological order and then it does not require spontaneous
symmetry breaking.
In Ref.~\cite{Kashiwa:2015tna}, the present authors consider the free-energy
degeneracy at finite $T$ based on the analogy of the ground-state degeneracy to
determine the deconfinement phase transition.
The modification of the topology is achieved by introducing the
dimensionless imaginary chemical potential, $\theta \equiv \mui/T \in
\mathbb{R}$ where
$\mui$ is the imaginary chemical potential.
As a result, the deconfinement transition temperature at zero real
quark chemical
potential ($\mur=0$) can be determined by the
Roberge-Weiss (RW) endpoint temperature, $T_\mathrm{RW}$.
The RW endpoint is the endpoint of the RW transition at $\theta=\pi/\nc$
where $\nc$ is the number of colors~\cite{Roberge:1986mm}.
This suggests that the deconfinement transition may be the topological
phase transition related with the topological order.

From differences of topology between the confined and deconfined phases
as mentioned above,
we can construct the quantum order-parameter which we call the quark
number holonomy~\cite{Kashiwa:2016vrl} as
\begin{align}
\Psi
&= \oint_0^{2 \pi}
   \mathrm{Im}
   \Bigl(
          \frac{d {\tilde n}_q}{d \theta}
   \Bigr) ~d \theta,
\label{Eq:hol}
\end{align}
where ${\tilde n}_q = n_q/T^3$ represents the quark number density
normalized by $T^3$ to make $\Psi$ dimensionless.
When $n_q$ has the gap at $\theta = \pi/\nc$, $\Psi$ should be nonzero
and it counts the number of the gapped points of $n_q$ along $\theta$;
\begin{align}
 \Psi
&= 2 \nc  \lim_{\epsilon \to 0}
   \mathrm{Im}~n_q \Bigl( \theta = \frac{\pi}{\nc} -\epsilon \Bigr).
 \label{Eq:Psi2}
\end{align}
Therefore, $\theta=\pi/\nc$ is an important and interesting point.
If the RW endpoint becomes the triple-point where three
first-order transition lines meet~\cite{D'Elia:2009qz,Bonati:2010gi}, the
point where $\Psi$ start to have nonzero value is shifted to lower
temperature than $T_\mathrm{RW}$.
This temperature is sometimes called
$T_\mathrm{Beard}$; see Ref.~\cite{Kashiwa:2016vrl} for details.
In both cases, we can see difference of topology between the confined
and deconfined phases and thus we can state that the deconfinement
transition is the topological phase transition.
It should be noted that the expression (\ref{Eq:hol}) is valid not only
at zero density but also finite density.

\begin{figure}[htbp]
 \centering
 \includegraphics[width=0.48\textwidth]{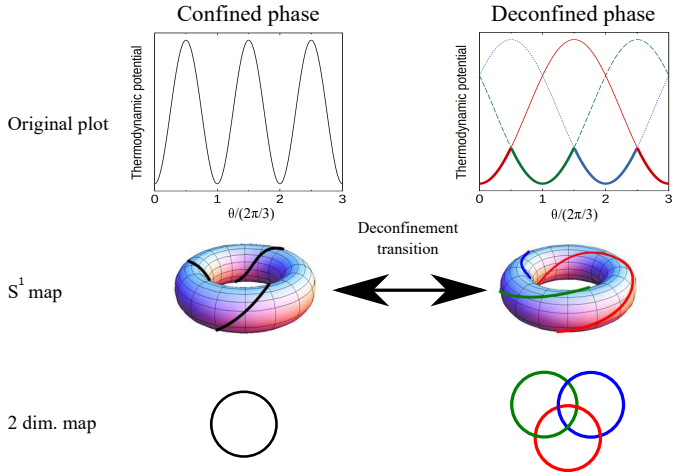}
 \caption{
 Schematic figures of the topological difference in the
 confined and deconfined phases with $\nc=3$.
The left (right) column represents the confined (deconfined) phase.
The first, second and third rows are the original $\theta$-dependence of
 the thermodynamic potential, its map to $S^1$ and its map to 2
 dimensional space by replacing the torus.
In the right panel of the first row, the bold line represents the global
 minima.
In the second row, solid lines which wind around torus are realized
 thermodynamic state
 (global minima of the effective potential) and its $\mathbb{Z}_3$
 images if those exist.
 }
\label{Fig:map}
\end{figure}

The topological structure can be visualized via
the map of thermodynamic quantities, for example the thermodynamic
potential, to the circle $S^1$ along periodic $\theta$ as shown in
Fig. \ref{Fig:map}.
Here, we plot the thermodynamic states which are global minimum and two
possible local minima of the thermodynamic potential at each $\theta$
as lines on the torus.
Two possible local minima are so called the $\mathbb{Z}_3$ images of the
global minimum.
In the confined phase, only one line winds around the torus $\nc$ times,
but there are $\nc$ lines in the deconfined phase and each line wind
around the torus only one time.
Number of winding are related with the RW
periodicity and its realization
mechanism~\cite{Roberge:1986mm,Kashiwa:2015tna}.
By removing the torus, lines can be mapped to the 2-dimensional space
as closed rings:
one ring exists in the confined phase and three entangled rings appear
in the deconfined phase.
These entangled rings are
not the Borromean rings because we cannot untangle them by removing
one ring.

To see more topological properties of thermodynamic states,
we consider three
operations, $\Theta$, $\Upsilon$ and $\mathbb{Z}_\mathrm{N_c}$, for
the thermodynamic state.
We set the global minimum of the thermodynamic
potential at $\theta=0$ as the initial thermodynamic state and
express it as $\psi(\theta=0,\phi=0)$ where $\phi$
represents the phase of the Polyakov loop.
In the case with $\mathrm{N}_c=3$, $\phi$ can take $0, 2\pi/3 $
and $4\pi/3$ in the deconfined phase.
Three operations are following:
($\Theta$ operation)
It shifts $\psi(\theta,\phi)$ to $\psi(\theta+2\pi,\phi')$ and then
we trace the state which thermodynamic potential decreases if states
intersect each other when $\theta$ is varying.
($\Upsilon$ operation)
It shifts $\psi(\theta,\phi)$ to $\psi(\theta+2\pi/\nc,\phi)$.
It can be interpreted that the certain flux is inserted to
the closed Euclidean temporal coordinate loop and then $\theta$ is
shifted by the Aharonov-Bohm effect.
($\mathbb{Z}_\mathrm{N_c}$ operation)
It is the standard $\mathbb{Z}_\mathrm{N_c}$
transformation which changes $\psi(\theta,\phi)$ to
$\psi(\theta,\phi+2\pi/\nc)$ if the $\mathbb{Z}_\mathrm{N_c}$ images
exist.
In the confined phase, these operations should commute with each other
because thermodynamic states at any $\theta$ belong to the same image
as shown in the left-top panel of Fig.~\ref{Fig:map}.
By comparison, three operations are not always commutable in the deconfined
phase and these show non-trivial commutation relations.

The topological determination of the deconfinement phase transition is
not yet applied to nonzero $\mur$ region and thus we investigate this
region in this paper by using the effective model approach.
At finite $\mur$, there is the sign problem in the lattice QCD
simulation and thus we cannot obtain reliable results.
Therefore, the effective model calculations are important
to have qualitative and quantitative understanding of the QCD phase
structure.
To describe the non-trivial free-energy degeneracy correctly,
effective models should possess several topological properties of QCD at
finite $\theta$.
Such effective models sometimes have the model sign
problem~\cite{Fukushima:2006uv,Tanizaki:2015pua} which, of course,
relates with the original sign problem.
The model sign problem can be resolved by using the Lefschetz-thimble
path-integral
method~\cite{Witten:2010cx,Cristoforetti:2012su,Fujii:2013sra} and then
we may investigate $\mur$ region directly~\cite{Tanizaki:2015pua},
but it is not an easy task at present.
We need some more technical extensions of
Ref.~\cite{Tanizaki:2015pua} in the complex $\mu$ case.
It should be noted that we can not use the standard approximation
for the model sign problem in the PNJL model
which only takes into account the real part of the thermodynamics
potential in the present case.
This approximation restricts $\Phi$ to only take real values.
At finite $\mui$, the phase of $\Phi$ plays a crucial role and thus
it is  necessary that $\Phi$ can take complex values.

In this paper, we consider $\muiso$
as an alternative approach to investigate the QCD phase structure at
finite $\mur$.
It is because phase diagrams at finite $\mur$ and $\muiso$ are
identical outside of the pion condensed region in the large-$\nc$ limit
via the well-known orbifold
equivalence~\cite{Cherman:2010jj,Cherman:2011mh,Hanada:2011ju}.

This paper is organized as follows.
In the next section, we explain the isospin chemical potential and
formulation of the effective model.
Numerical results are shown in Sec.~\ref{Sec:Num}
Section~\ref{Sec:Summary} is devoted to summary.

\section{Effective model with isospin chemical potential}
\label{Sec:Iso}
In this paper, we consider $\muiso$ to extract
information of the $\mur$ region because the $\muiso$ region has
following two characteristic properties:
\begin{description}
 \item[1. Sign problem free]\mbox{}\\
	    It is well known that there is no sign problem because the
	    condition $ \tau_2
	    \gamma_5 {\cal D} \gamma_5 \tau_2 = {\cal D}^\dag$ is
	    manifested where ${\cal D}$ denotes the Dirac operator and
	    the symbol $\tau_i$ mean Pauli matrices in the flavor
	    space. This leads the condition $\mathrm{det}({\cal D}) \ge 0$
	    even if $\mui$ is nonzero.
 \item[2. Orbifold equivalence]\mbox{}\\
	    The phase structure at finite $\muiso$ is identical
	    with that at finite $\mur$ outside of the
	    pion-condensed phase at least in the large-$\nc$
	    limit~\cite{Hanada:2011ju}.
	    This equivalence is violated with finite $\nc$ by $1/\nc$
	    corrections such as the flavor mixing loops.
\end{description}
With the two properties, we can expect that the qualitative information
of QCD phase diagram at finite $\mur$ can be extracted from the QCD
phase diagram at finite $\muiso$.
For example, the no-go theorem of critical phenomena were obtained in
Ref.~\cite{Hidaka:2011jj} via the orbifold equivalence.
The mean-field approximation which picks up
the leading-order of the $1/\nc$ expansion is widely used in the effective
model calculation to investigate the QCD phase structure and thus our
analysis based on the orbifold equivalence can be acceptable for a first
step of the issue to understand the topological deconfinement transition
at finite $\mur$.
Therefore, our results may provide positive motivation to perform
the lattice QCD simulation in the
$\muiso \in \mathbb{R}$ region with finite $\mui$.
It should be noted that the sign problem comes back if we
consider the different $\theta$ for two quark fields with different
flavor.
In this case, the imaginary part of $\muiso$ appears in addition to
its real part and then the condition,
$ \tau_2 \gamma_5 {\cal D} \gamma_5 \tau_2 = {\cal D}^\dag$, is no
longer valid.
This issue does not affect our present discussions, but it should be
considered when we take into account the imaginary part of $\muiso$ in the
future.

To investigate the phase structure at finite $T$ and $\mu_\mathrm{iso}$,
we use the Polyakov-loop extended Nambu--Jona-Lasinio (PNJL)
model~\cite{Fukushima:2003fw}.
This model can well reproduce the
QCD properties at finite $\theta$; see
Ref.~\cite{Sakai:2008py} as an example.
The Lagrangian density of the two-flavor and three-color PNJL model in
the Euclidean space is
\begin{align}
 {\cal L}
 &= {\bar q} (\Slash{D} + m_0) - G[({\bar q}q)^2+({\bar
 q}i\gamma_5 {\vec \tau} q)^2]
+ {\cal V}_{\mathrm{g}} (\Phi,{\bar \Phi}),
\end{align}
where $m_0$ means the current quark mass, the covariant derivative is
$D_\nu - i g A_\nu \delta_{\nu 4}$,
$\Phi$ (${\bar \Phi}$) denotes the Polyakov-loop (its conjugate) and
${\cal V}_{\mathrm{g}}$ expresses the gluonic contribution.
In the case with $m_0=\muiso=0$, the Lagrangian density has
$SU(2)_L \times SU(2)_R \times U(1)_V \times SU(3)_\mathrm{color}$ symmetry.
With $m_0 \neq 0$ and $\mui \neq 0$, the Lagrangian density has
$U(1)_{{\bf I}_3} \times U(1)_V \times SU(3)_\mathrm{color}$ symmetry.
For example, see
Refs.~\cite{Endrodi:2014lja,Brandt:2016zdy,Brauner:2016lkh,Cao:2016ats}
for recent progress in the investigation of QCD phase structure at
finite $\muiso$.

The imaginary and isospin chemical potentials can be introduced to the
model as the form
\begin{align}
\mu = \mu_q {\bf I} + \mu_\mathrm{iso} \tau_3,
\end{align}
where ${\bf I}$ is the unit matrix.
Each $\mu_q$ and $\muiso$ can be represented as
\begin{align}
\mu_q = \frac{\mu_u + \mu_d}{2},~~~~
\muiso = \frac{\mu_u-\mu_d}{2},
\end{align}
where $\mu_u$ ($\mu_d$) is the quark chemical potential of up (down)
quark.
In this paper, we consider $\mu_q = i \mui$ ($\mui \in \mathbb{R}$)
and $\muiso \in \mathbb{R}$ to discuss the topological
deconfinement transition at finite $\muiso$.

With the mean-field approximation,
the thermodynamic potential can be expressed as
\begin{align}
{\cal V}
&= {\cal V}_{\mathrm{f}} + {\cal V}_{\mathrm{g}},
\end{align}
where ${\cal V}_\mathrm{f,g}$ are the fermionic and gluonic parts,
respectively.
The actual form of ${\cal V}_\mathrm{f}$ is
\begin{align}
{\cal V}_\mathrm{f}
&= - 2 \sum_{i=\pm} \int \frac{d^3 p}{(2\pi)^3}
   \Bigl[ 3 E_{i,{\bf p}}
        + T \ln \Bigl( f_i^- f_i^+ \Bigr)\Bigr]
\nonumber\\
&+ G (\sigma^2 + \pi^2),
\end{align}
where
\begin{align}
f^-_i
&= 1
 + 3 (\Phi+{\bar \Phi} e^{-\beta E_{i,{\bf p}}^-} ) e^{-\beta E_{i,{\bf p}}^-}
 + e^{-3\beta E_{i,{\bf p}}^-},
\nonumber\\
f^+_i
&= 1
 + 3 ({\bar \Phi}+\Phi e^{-\beta E_{i,{\bf p}}^+} ) e^{-\beta E_{i,{\bf p}}^+}
 + e^{-3\beta E_{i,{\bf p}}^+}.
\end{align}
Quark single particle energies are given as
$E_{\pm{\bf p}}^\mp = E_\pm \mp i\mui$ with
$E_\pm = \sqrt{(E_{\bf p} \pm \muiso)^2 + N^2}$ for $E_{\bf p} =
\sqrt{{\bf p}^2 + M^2}$.
Symbols $M$ and $N$ are defined as $M=m_0 - 2G\sigma$ and $N=-2G\pi$
with $\sigma = \langle {\bar q} q \rangle$ and
$\pi = \langle {\bar q} i \gamma_5 \tau_1 q \rangle$.
Here we take the order parameter
of the $U(1)_{{\bf I}_3}$ symmetry breaking as
\begin{align}
 \pi^\pm = \frac{1}{\sqrt{2}} e^{\pm i\varphi} = \langle {\bar q} i
 \gamma_5 \tau_\pm q \rangle,
\end{align}
and we can set $\varphi=0$ as in
Ref.~\cite{He:2005nk,Zhang:2006gu} without loss
of generality.
In the following disucssions, we set
$\pi = |\langle {\bar q} i \gamma_5 \tau_1 q \rangle|$.
For example, see Ref.~\cite{He:2005nk,Sun:2007fc,He:2006tn} for
discussions in the isospin chemical potential region by using the NJL
model.

We employ the Polyakov loop potential used in
Ref.~\cite{Roessner:2006xn} as
${\cal V}_{\mathrm{g}}$ and the parameter set used in
Ref.~\cite{Kouno:2011zu};
for details of the PNJL model with $\mu_\mathrm{iso}$, see
Refs.~\cite{Zhang:2006gu,Mukherjee:2006hq,Kouno:2011zu} as an example.
In the PNJL model, the {\it model Polyakov-loop} is defined as
\begin{align}
\Phi
&= \frac{1}{\nc} \mathrm{tr}_c
   \Bigl[e^{i g \langle A_4 \rangle / T} \Bigr],
\label{Eq:Pol}
\end{align}
with the Polyakov-gauge fixing.
The Polyakov loop is, of course, gauge invariant, but
Eq. \eqref{Eq:Pol} is not;
see Ref.~\cite{Braun:2007bx} for discussions on the model
and correct Polyakov-loops from the Jensen inequality.
Nevertheless, the model Polyakov loop Eq.~\eqref{Eq:Pol} is enough for
our purpose.
The Polyakov loop is introduced to just
model the $\mathbb{Z}_3$ properties of QCD because
the modeling of $\mathbb{Z}_3$ properties plays a crucial role
to reproduce characteristic properties of QCD at finite $\theta$.
Thus, there is no need to reproduces the
correct Polyakov-loop by the model Polyakov-loop.
It is the reason why we use the simplest PNJL model in this paper.

The PNJL model seems to be an adequate effective model to describe
the topological properties of QCD at finite $\mui$ and thus it allows us
to discuss the topological confinement-deconfinement transition.
Our numerical results presented in the next section are performed in the
situation that the
topological confinement-deconfinement transition is modeled at $\mur=0$
and after the $\muiso$-dependence of the transition is discussed.
It should be noted that we do not claim that the PNJL model can
reproduce all of QCD properties.

\section{Numerical results}
\label{Sec:Num}

\begin{figure}[htbp]
 \includegraphics[width=0.45\textwidth]{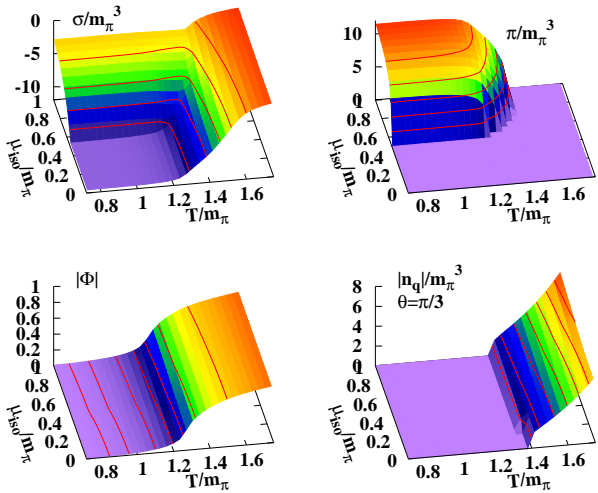}
 \caption{The $T$ and $\muiso$ dependence of the chiral condensate, the
 pion condensate and the Polyakov loop at $\theta=0$ and the quark
 number density at $\theta=\pi/3$.
 Quantities are normalized by $m_\pi=138$ [MeV].}
\label{Fig:T-muiso}
\end{figure}

To investigate the topological deconfinement transition,
$\theta=\pi/3$ region is important as shown in Eq.~\eqref{Eq:Psi2} and
thus we show the quark number density at $\theta = \pi/3$ in addition to
the chiral condensate, the pion condensate and the Polyakov loop at
$\theta=0$ here.
Figure~\ref{Fig:T-muiso} shows $|\sigma|/m_\pi^3$, $\pi/m_\pi^3$,
$|\Phi|$ and $|n_q|/m_\pi^3$ as a function of $T/m_\pi$ and $\muiso/m_\pi$.
The chiral and pion condensates strongly depend on $\muiso$ as well as
on $T$, but the Polyakov loop and the quark number
density are less sensitive to $\muiso$.
In the leading-order approximation of the PNJL model, the gluonic
contributions do not have the explicit chemical-potential dependence because
there is no quark polarization effects which are higher order
contributions of the $1/\nc$ expansion.
Thus, $\muiso$ cannot affect the Polyakov loop directly.
It is interesting to compare these results with the
isospin-density dependence of the Polyakov-loop on the
lattice~\cite{Endrodi:2014lja}.
Interesting point is that the quark number density is also insensitive
to $\muiso$ and it means that the chiral and deconfinement transition are
not correlated strongly.
This tendency can be expected to appear at finite $\mur$ via the orbifold
equivalence outside of the pion-condensed phase.

\begin{figure}[h]
 \includegraphics[width=0.23\textwidth]{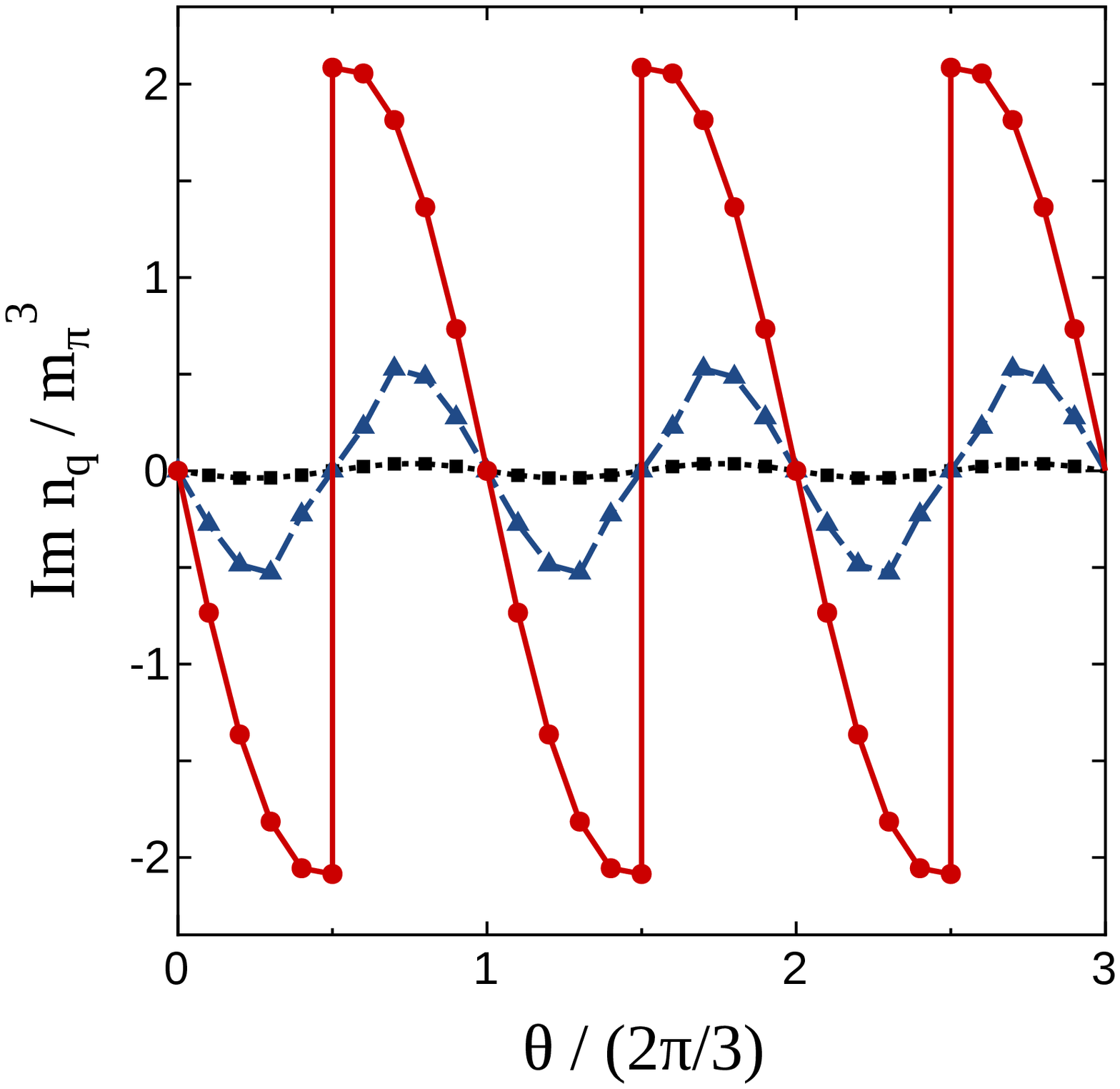}
 \includegraphics[width=0.23\textwidth]{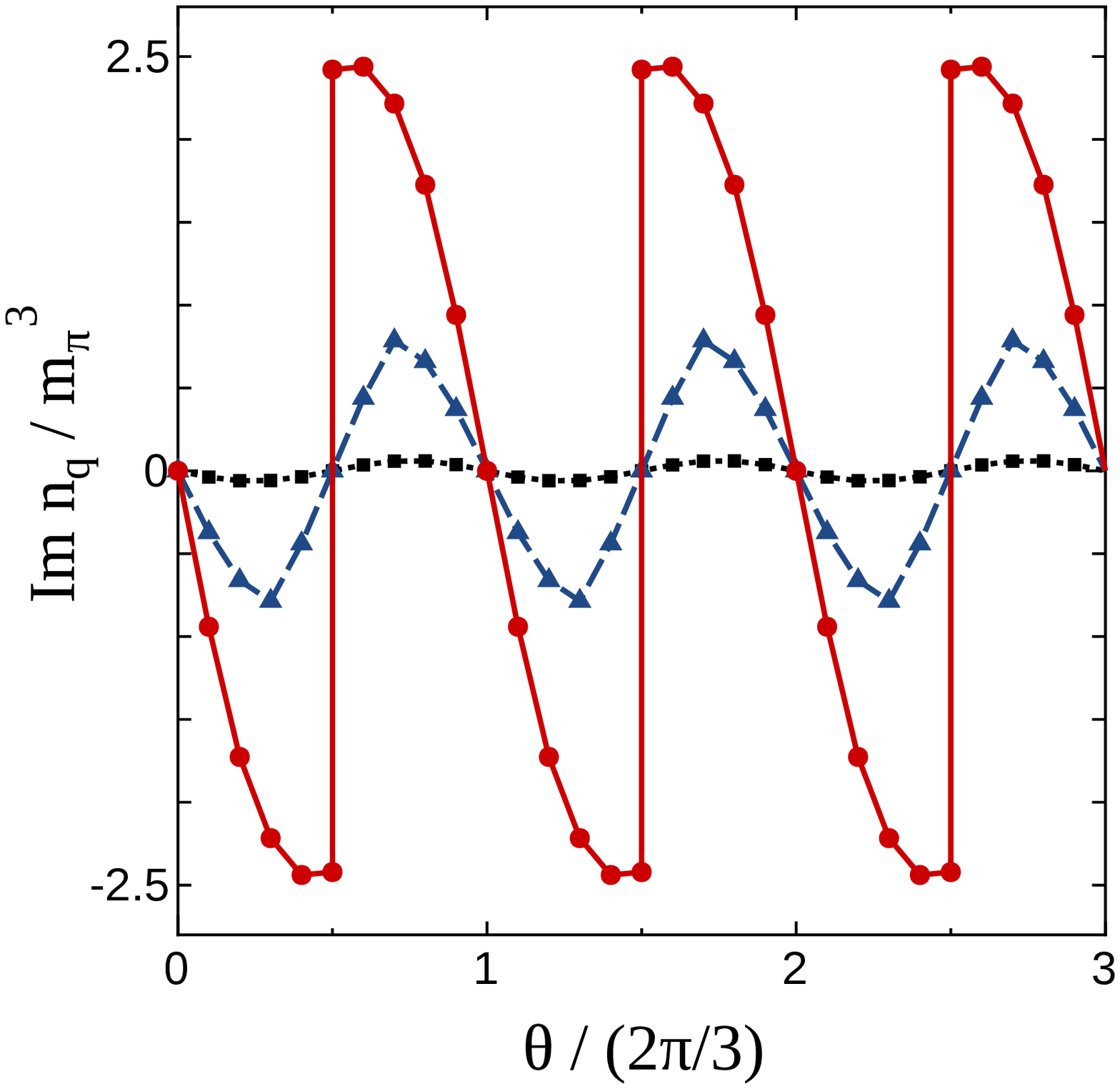}
 \caption{ The $\theta$ dependence of the quark number density
 with $\muiso = 0$ and $0.8 m _\pi$.
 The dotted, dashed and solid lines are results at $T=160$, $180$ and
 $200$ MeV, respectively.
 }
\label{Fig:nq}
 \end{figure}

\begin{figure}[htbp]
 \includegraphics[width=0.38\textwidth]{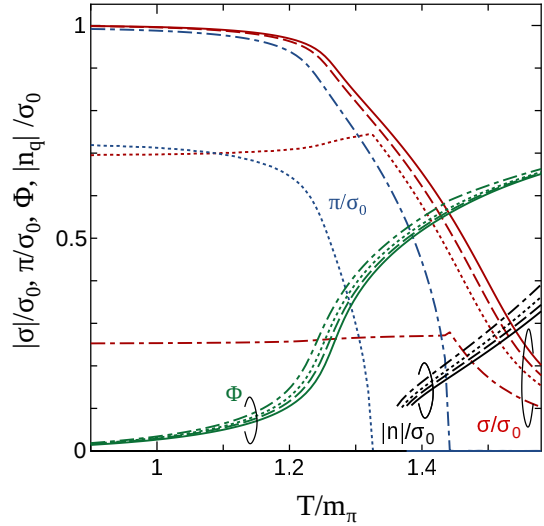}
 \caption{The $T$ dependence of the chiral condensate, the pion
 condensate and the
 Polyakov loop at $\theta=0$ and the quark number density at $\theta=\pi/3$.
 The solid, dashed, dotted, and dot-dashed lines are these quantities
 at $\mu_\mathrm{iso}/m_\pi=0, 0.4, 0.6, \mathrm{and}\ 1$, respectively.
 Quantities are normalized by
 $\sigma_0=|\sigma_\mathrm{vacuum}|$.
 The step-size of $T$ is 1 [MeV].
 }
\label{Fig:Tdep}
\end{figure}

In the topological determination of the confinement-deconfinement
transition, the behavior of quark number density at finite $\theta$ is
important.
The left (right) panel of Fig.~\ref{Fig:nq} shows the behavior of $n_q$
as a function of $\theta$ with $\muiso =0$ ($0.8m_\pi$) for $T=160$,
$180$ and $200$ MeV.
The $\theta$ dependence of the quark number density clearly shows
the general feature of the topological phase transition discussed in
Sec. 1: It is smooth below $T_\mathrm{RW}$ and has gaps above
$T_\mathrm{RW}$.
By comparison, the quark number density does not strongly depend on
$\muiso$.
This behavior means that the topological confinement-deconfinement
temperature is not very sensitive to $\muiso$ in the present model.
For reader's convenience, we also show the
$T/m_\pi$ dependence of $|\sigma|/\sigma_0$, $\pi/\sigma_0$ and $\Phi$
at $\theta=0$ and $|n_q|/\sigma_0$ at $\theta=\pi/3$ with $\muiso/m_\pi=0,
0.4, 0.6, 1$ in Fig.~\ref{Fig:Tdep}.

\begin{figure}[h]
 \centering
 \includegraphics[width=0.38\textwidth]{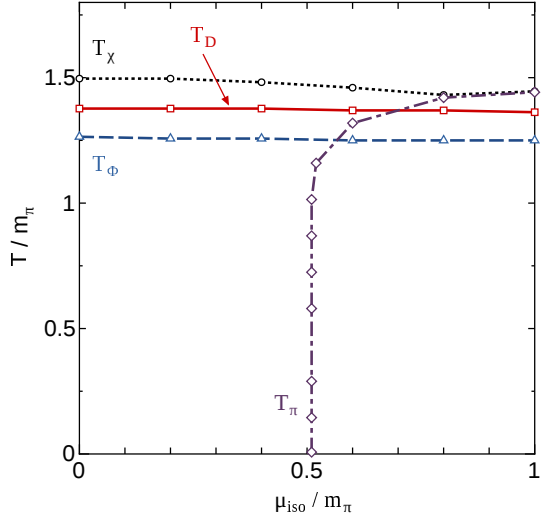}
 \caption{ The phase diagram in the $T$-$\mu_\mathrm{iso}$ plane.
 Symbols are numerical results and lines are linear interpolations
 between them.
 The solid, dashed, dotted and dot-dashed lines are explained in the
 text.
 Numerical error of these lines are about $\pm 1.5$ [MeV].
 }
\label{Fig:PD}
\end{figure}

The phase diagram in the $(T,\muiso)$ plane is shown
in Fig.~\ref{Fig:PD}.
The solid, dashed, dotted and dot-dashed lines are the deconfinement
transition temperature ($T_\mathrm{D}$),
the deconfinement pseudo-critical temperature determined by the peak
of $d\Phi/dT$ ($T_\Phi$),
the chiral pseudo-critical temperature determined by the peak of
$d\sigma/dT$ ($T_\chi$)
and the phase transition temperature of the pion condensation ($T_\pi$).
If $d \sigma / dT$ and $d \Phi / dT$ have two peaks, we choice higher one as
the pseudo-critical temperature.
We determine $T_\mathrm{D}$ by $T_\mathrm{RW}$ in this paper because
the triple-point nature of the RW endpoint depends on the form of
${\cal V}_\mathrm{g}$ in the PNJL model and also
the qualitative behavior of $T_\mathrm{RW}$ and $T_\mathrm{Beard}$ are
basically same.
From Fig.~\ref{Fig:PD}, we can find following features:
\begin{description}
 \item[1. Behavior of $\mathbf{T_D}$ and $\mathbf{T_\Phi}$]\mbox{}\\
	    $T_\mathrm{D}$ and $T_\Phi$ slightly decrease with increasing
	    $\muiso$ and show similar qualitative behaviors.
	    However, $\Phi$ is continuously changed at finite $T$ and
	    thus the determination of $T_\Phi$ is not clear unlike
	    $T_\mathrm{D}$.
 \item[2. Behavior of $\mathbf{T_D}$ at finite density]\mbox{}\\
	    $T_\mathrm{D}$ decreases with increasing $\mu_\mathrm{iso}$
	    and thus we can expect that $T_\mathrm{D}$ also show
	    decreasing behavior at finite $\mur$ via the orbifold
	    equivalence. This result is
	    consistent with our perturbative estimate of $T_D$ at small
	    $\mur$~\cite{Kashiwa:2015tna}.
 \item[3. Relation between $\mathbf{T_D}$ and $\mathbf{T_\chi}$] \mbox{}\\
	    The present definition of $T_D$ is not affected much by the
	    chiral transition even though we use the quark-bilinear
	    $\langle {q^\dag q} \rangle$.
 \item[4. Relation between $\mathbf{T_D}$ and $\mathbf{T_\pi}$]
            \mbox{}\\
	    By comparison, $T_D$ seems to be smaller than $T_\pi$ at
	    large $\mu_\mathrm{iso}$.
	    This is not unreasonable since these transition are
	    independent of each other.
	    Finite quark mass leaves $\sigma$ finite and pion keeps its
	    Nambu-Goldstone boson nature even at large $T$.
	    For example,
	    some effective model results including the present one
            show $T_\chi > T_\mathrm{D}$ at $\mur=0$.
\end{description}
We also note that $T_D$ is smaller than $T_\chi$ at small $\muiso$.
When the pseudo-critical chiral transition boundary ends at the RW endpoint,
one should find $T_\chi(\theta=0)<T_\chi(\theta=\pi/3)=T_\mathrm{RW}(=T_D)$.
It should be noted that this is not always the case;
the order of two temperatures, $T_\chi$ and $T_\mathrm{RW}$,
depends on the details of the framework.
In the present setup,
the chiral transition boundary reaches the RW transition line ($\theta=\pi/3$)
at significantly higher temperature than the RW endpoint,
$T_\chi(\theta=\pi/3)>T_\mathrm{RW}$.
It should be noted that the difference between the
confinement-deconfinement temperatures determined from $T_\mathrm{RW}$
and $T_\mathrm{Beard}$ are almost the same in the present model.
We check it at $\mu_\mathrm{iso} = 0$ and $0.8 m_\pi$.
Actually, the difference,
$(T_\mathrm{RW}-T_\mathrm{Beard})/T_\mathrm{RW}$, is less than $3$\% and
the slightly decreasing behavior of $T_\mathrm{Beard}$ with increasing
$\mu_\mathrm{iso}$ also appears.

Sensitivity of the critical temperature on the chemical potential
is quantified using the curvature $\kappa$,
\begin{align}
\frac{T_c(\mu)}{T_c(\mu=0)}
=1 - \kappa\,\left(\frac{\nc\mu}{T}\right)^2 + \mathcal{O}\left[\left(\frac{\nc\mu}{T}\right)^4\right]
\ .
\end{align}
For $T_c=T_\chi, T_\Phi$ and $T_\mathrm{D}$ with $\mu=\muiso$,
the curvatures are obtained as
$\kappa_\chi=0.017\pm0.001$,
$\kappa_\Phi=0.004\pm0.001$
and
$\kappa_\mathrm{D}=0.003\pm0.001$,
respectively.
Compared with the chiral transition,
curvatures in $T_\Phi$ and $T_D$ are much smaller,
and show that the deconfinement temperature is much less sensitive
to $\muiso$ than the chiral transition temperature.
The chiral transition curvature is larger than the lattice result
in the real quark chemical potential,
$\kappa \simeq 0.008$~\cite{Allton:2002zi},
but it is within a two sigma uncertainty of more recent lattice results
$\kappa=0.013\pm0.003$~\cite{Bonati:2014rfa},
$\kappa=0.0135\pm0.002$~\cite{Bonati:2015bha},
$\kappa=0.0149\pm0.0021$~\cite{Bellwied:2015rza} and
$\kappa=0.020\pm0.004$~\cite{Cea:2015cya}.
The agreement would be accidental,
but suggests the validity of the orbifold equivalence.

\section{Summary}
\label{Sec:Summary}
In this paper, we have discussed the deconfinement transition from
topological viewpoints.
Modification of the topology of the system is achieved by introducing
the dimensionless imaginary chemical potential ($\theta$).

The topological difference between the confined and deconfined is
visualized by using the map of the thermodynamic quantities to circle
$S^1$ along periodic $\theta$.
Then, we find that there are three entangled rings of thermodynamic states
in the deconfined phase, but there is only one ring in the confined
phase after performing the $2$-dimensional map of thermodynamic states.
To understand it more clearly, we have considered three operations for
the thermodynamic state,
$\theta$ shift, the flux insertion to the closed Euclidean temporal
coordinate loop
and $\mathbb{Z}_\mathrm{N_c}$ transformation,
and found nontrivial commutation relations between them in the
deconfined phase.

To investigate the topological deconfinement phase transition in the
real chemical potential ($\mur$) region, we have investigated the isospin
chemical potential ($\muiso$) region. Phase diagrams in
both regions are identical to each other outside of the pion-condensed
phase via the orbifold equivalence at least in the large-$\nc$ limit.
The chiral and pion condensates are strongly affected by
$\muiso$, but the Polyakov-loop and the quark number density at $\theta
= \pi/3$ are not.
We show this feature by estimating the curvature of transition
lines.
In the present PNJL model treatment, the gluonic part does not have
explicit chemical potential dependence because the quark polarization
effects are ignored.
This leads to uncorrelated results between the chiral and
deconfinement transitions.
Some results for the chiral and pion condensations have been
observed on the
lattice~\cite{Kogut:2002tm,Kogut:2002zg,Kogut:2004zg,deForcrand:2007uz,Detmold:2012wc,Endrodi:2014lja,Brandt:2016zdy},
but not for the quark number density at $\theta=\pi/3$.
If we will access lattice QCD data of the
quark number density at finite $\muiso$ with $\theta=\pi/3$ in the
future, we can judge our surmise.
If lattice simulations will provide the strong correlated results
between the chiral and the deconfinement transitions at finite $\muiso$,
we can use the $\mu_\mathrm{iso}$ region as a laboratory
to extend the gluonic part of the effective
model to include quark polarization effects by comparing the deconfinement
transition in the effective model and that in lattice
QCD simulations.

In this study, we clarify topological properties of QCD via the
imaginary chemical potential.
Also, it is a first study of the topological deconfinement transition
at finite chemical potential.
We hope that this study shed a light to nature of the deconfinement
transition from topological viewpoint.
One of the possible future directions of this study is the
calculation of
the Uhlmann phase~\cite{Uhlmann1986} which has been used to
discuss the finite temperature topological-order in the condensed matter
physics~\cite{Viyuela2014,viyuela2014two}.
Therefore, we can clarify the confinement-deconfinement transition from
the same context of the
condensed matter physics if we can calculate the Uhlmann phase in QCD.
The calculation of
the Uhlmann phase in QCD seems to be very difficult or impossible at
present, but it will complete our discussions presented in Refs.
\cite{Kashiwa:2015tna,Kashiwa:2016vrl} and also this paper.

The authors thank H. Kouno for his useful comments.
K.K. is supported by Grant-in-Aid for Japan Society for the Promotion
 of Science (JSPS) fellows No.26-1717.
A.O. is supported in part by the Grants-in-Aid for Scientific Research
 from JSPS (Nos. 15K05079, 15H03663, 16K05350),
the Grants-in-Aid for Scientific
 Research on Innovative Areas from MEXT (Nos. 24105001, 24105008),
 and by the Yukawa International Program for Quark-hadron
 Sciences (YIPQS).

\bibliographystyle{elsarticle-num}
\bibliography{ref}

\end{document}